\documentclass[11pt,a4paper]{article}

\usepackage[top=2.5cm, left=2.5cm, right=2.5cm, bottom=2.5cm]{geometry}

\usepackage[utf8]{inputenc}
\usepackage{amsmath}
\usepackage{graphicx}
\usepackage{booktabs}
\usepackage{mathptmx}
\usepackage[normalem]{ulem}
\useunder{\uline}{\ul}{}
\usepackage{ntheorem}
\usepackage[toc,page]{appendix}
\usepackage{ntheorem}
\usepackage{tabularx}
\usepackage{multicol}
\usepackage{multirow}
\usepackage{verbatim,float}
\usepackage{longtable}
\usepackage{tikz}
\usepackage[numbers]{natbib}
\usepackage{times}
\usepackage{float}
\usepackage{enumerate}
\usepackage{romannum}
\usepackage{xcolor}
\usepackage[colorlinks=true,linkcolor=black]{hyperref}
\hypersetup{colorlinks = true, linkcolor = blue, anchorcolor = blue, citecolor = blue, filecolor = blue, urlcolor = blue}

\definecolor{TextColA}{rgb}{1,0,0}

\usepackage{color, colortbl}

\title{Automated lag-selection for multi-step univariate time series forecast using Bayesian Optimization: Forecast station-wise monthly rainfall of nine divisional cities of Bangladesh} % The article title

\author{Rezoanoor Rahman$^1$ and Fariha Taskin$^{1,*}$ } 
\date{
	\footnotesize{1. Institute of Statistical Research and Training, University of Dhaka, Bangladesh}\\
\footnotesize{$^*$Corresponding Author: iftakhar@isrt.ac.bd} } 

\begin{document}
	
%	\maketitle
	\pagenumbering{arabic}

	\begin{center}
		\large{{\bf Automated lag-selection for multi-step univariate time series forecast using Bayesian Optimization: Forecast station-wise monthly rainfall of nine divisional cities of Bangladesh}}\\
		\vspace{.3in}

		%M. Iftakhar Alam, % Barbara Bogacka$^{2}$ and D. Stephen Coad$^{2}$\\
		
		Rezoanoor Rahman$^1$ and Fariha Taskin$^{1}$ \\
%		\vspace{.5cm}
		\footnotesize{$^1$Institute of Statistical Research and Training, University of Dhaka, Dhaka 1000, Bangladesh}\\
%		\footnotesize{$^*$Corresponding Author: iftakhar@isrt.ac.bd, Phone: +8801716187606}\\
%		\footnotesize{2. School of Mathematical Sciences, Queen Mary, University of London, London E1 4NS, U.K.}\\
		\vspace{1cm}
		Abstract
	\end{center}
	%
	%\vspace{1cm}

Rainfall is an essential hydrological component and most of the economic activities of an agrarian country like Bangladesh depend on rainfall. An accurate rainfall forecast can help make necessary decisions and reduce damages caused by heavy or low or no rainfall. The monthly average rainfall is a time series data, and recently, long-short-term memory (LSTM) neural networks are being used heavily for time series forecasting problems. One major challenge of forecasting using LSTMs is to select the appropriate number of lag values. In this research, we considered the number of lag values selected as a hyperparameter of LSTM; it, together with the other hyperparameters determining the structure of LSTM, was optimized using Bayesian optimization. We used our proposed method to forecast rain for nine different weather stations in Bangladesh. Finally, the performance of the proposed model has been compared with some other LSTM with different lag-selection methods and several popular machine learning and statistical forecasting models.\\
		

%%%%%%%%%%%%%%%%%%%%%%%%%%%%%%%%%%%	
\section{ Introduction}
\label{intro}

Bangladesh has a largely agrarian economy; its agriculture comprises about 18.6 percent of its GDP and employs around $45$ percent of the total labor force \cite{agriculture_bd}. Rainfall is a central element of the hydrological process and a significant portion of the agricultural production system is dependent on rainfall water through irrigation. Due to this dependency, a significant deviation in rainfall can cause severe damage to the economy. The damage is twofold. In one extreme, if the rainfall is inadequate or highly irregular, it can cause drought \cite{drought_3}. Bangladesh faces a significant drought every 2.5 years \cite{drought_2}, which is the main drawback to ensuring food security in Bangladesh despite significant improvements in food production because drought affects the agricultural sector in Bangladesh during rice production, which is the main crop in Bangladesh \cite{drought_1}. In another extreme, due to excessive rainfall, floods can occur because Bangladesh geographically has a unique setting for flooding. About 80 percent of the landmass of Bangladesh is plain, while most parts of the country are low-lying, and as a result, it is highly vulnerable to the threat of repeated floods. From $1971-2014$, 6.31 million people become victims of floods yearly on average, and the total monetary value of the damage was $12232.70$ million USD \cite{flood_affects_bd}. Another problem becoming a primary concern in Bangladesh, especially at urban areas, is water logging caused by heavy rainfall. A recent study shows that 76 percent of traffic movement gets disrupted, 82 percent of the roads get damaged, and 68 percent of the water gets polluted in the waterlogged areas \cite{water_logging_new}.

As the losses are because of the events caused by deviations in rainfall, it is necessary to forecast rainfall accurately. Additionally, accurate rainfall can help policymakers to make essential decisions regarding different agricultural policies, especially for a country like Bangladesh. However, there has not been much research on forecasting rainfall more accurately. Among the available research, studies by Mahmud et al. \cite{rainfall_bd_1}, and Mahsin et al. \cite{rainfall_bd_2} are notable. 

%However, they used ARIMA methods for forecasting the rainfall in Bangladesh, which has been found to perform worse than many machine learning and neural network-based forecasting algorithms in the literature.

%%Moreover, because of these serious risks, rainfall should be forecasted at a time that is reasonably before the events when an extreme rainfall event can occur to take necessary steps for mitigating or preventing the losses. 

In contrast to Bangladesh, different time series forecasting methods have been used for forecasting rainfall in the literature. Many statistical, machine learning (ML), and neural network (NN) based models have been proposed in the literature to forecast rainfall in different parts of the world. Among the statistical methods, seasonal autoregressive integrated moving average (SARIMA) is the most commonly used \cite{arima_extra_1, arima_extra_2} because of its capability of producing both point and interval forecasts and the well-defined Box-Jenkins method for selecting the best model \cite{box_jenkins}. ML models like support vector regression (SVR) and random forest (RF) have also been implemented successfully for rainfall forecasting \cite{rainfall_rf_svm}. Among the NN models for forecasting rainfall, the use of single hidden-layer artificial NN models \cite{indian_rainfall_1,rainfall_ann}, convolutional neural network (CNN) models are noticeable \cite{rainfall_cnn}. Recurrent neural network (RNN)\cite{rnn_intro} models can capture the temporal relationship between different inputs and, as a result, are being used for time-series forecasting problems \cite{rnn_1, rnn_3}. However, RNN faces vanishing gradient problem \cite{vanish_gradient_main} and, as a result, fails to capture the long-term time dependence in time series. In this regard, long short-term memory (LSTM) neural networks \cite{lstm_into} are particularly helpful for time series forecasting. Recently, LSTM has been extensively used for forecasting rainfall \cite{rainfall_lstm} as well as many other time series problems like demand forecasting \cite{demand_forecasting}, wind power forecasting, and many more \cite{lstm_multiple_input}. In much experimental research, LSTM has been shown to perform better than other neural network structures for time series forecasting \cite{lstm_better_ts_1,demand_forecasting}.

One of the major challenges of using a NN-based model is to find the best set of hyperparameters because the performance of a neural network model is highly dependent on its hyperparameters. The number of hyperparameters in a NN model is higher than in other ML models like SVM or RF. The most commonly used method is searching for hyperparameters manually. Nevertheless, it becomes inconvenient when the number of hyperparameters is enormous, as it is pretty easy for someone to misinterpret the trends of the hyperparameters \cite{human_problem}. As a result, using automatic search algorithms is necessary. Grid search, an automatic search algorithm that trains a ML model with each combination of possible values of hyperparameters on the training set and outputs hyperparameters that produce the best possible result, was proposed \cite{book_tibshirani}. However, as the number of hyperparameters increases, so does the number of combinations. Thus, grid search becomes inconvenient for computationally costly functions like LSTM models, which contain many hyperparameters. Random search, another automatic search algorithm that searches random combinations of hyperparameters within a range of values, is more feasible computationally but unreliable for training complex models \cite{random_search}. In this regard, Bayesian optimization (BO) combines prior information about the unknown function with sample information to obtain posterior information on the function distribution using the Bayes formula. Based on this posterior information, BO decides where the function obtains the optimal value. In many experiments, BO has been found to outperform other global optimization algorithms \cite{bayesian_better}. Recently, BO has been extensively used for finding the best set of hyperparameters while using NN models for sequential data \cite{bayesian_time}.

The inputs for forecasting time series models using ML or NN-based models are simply past observations. The selection of the number of past observations or LAGS significantly affects the forecasts. Selecting LAGS too small might be inadequate to learn the pattern of the time series data; on the other hand, selecting LAGS more than needed can make the model unnecessarily complex and overfit the model and hence produce bad forecasts. The most common approach for selecting LAGS is to consider it as a hyperparameter and search manually \cite{demand_forecasting}. The problem with this approach is that if the number of hyperparameters considered is large, considering LAGS as another hyperparameter, multiply the possible number of combinations several times. Another common manual approach to select this is to analyze the partial autocorrelation function (PACF) values and stop when the plot starts diminishing \cite{beta_pso}. One problem with this approach is that the PACF generally explains the linear correlation between the values and cannot capture non-linear trends in the series, while most machine learning and deep learning-based models are not linear. Another approach is to fix a particular set of hyperparameters, monitor the loss function, and search for LAGS, after which the loss function gets flattened. The limitation of this approach is that there is no guarantee that an optimal value of lags for one set of hyperparameters is optimal for another set of hyperparameters. Among the automated methods, selecting LAGS equal to the seasonal length is standard. Another approach is to select it equal to the length of the forecasting horizon. Moreover, Hewamalage et al. used a heuristic approach to multiply the forecast horizon or seasonal period by 1.25 and use this value in the selected LAGS \cite{recurrent_all}. 

Note that none of these automated methods consider LAGS as a hyperparameter. In this research, we considered LAGS as a hyperparameter in addition to the hyperparameters associated with the structure of the LSTM; thus, we automated the process of selecting Lag values using BO. Then we applied our proposed method to forecast total rainfall data for nine divisional cities of Bangladesh: Barisal, Dhaka, Comilla, Mymensingh, Rangpur, Chittagong, Rajshahi, Sylhet, and Khulna. To compare the performance of our proposed model, two popular machine learning models, RF and SVR, and two popular statistical time series forecasting techniques, SARIMA and exponential smoothing (ETS), are used.

%%%%%%%%%%%%%%%%%%%%%%%%%%%%%%%%%%%	

\section{ Methodology}
\label{method}

\subsection{Overview of Recurrent neural network and Long Short-Term Memory network}

The recurrent neural network (RNN) is a bioinspired neural network model \cite{rnn_intro} that uses the output of any stage as an input of the next step and thus captures the temporal relationship between inputs and, as a result, is widely used for modeling where a temporal relationship is present such as time series forecasting problems. For comparing RNN with other Neural Network structures, for producing the output of a particular state, the RNN considers information from the previous state in addition to the information from the inputs. In general, the weights of the RNN model are optimized using back-propagation through time (BTT) \cite{btt}. The major limitation of the RNN model is that it suffers from the vanishing gradient problem and, as a result, faces information loss in case of long-term dependence \cite{vanish_gradient_main}.

The Long short-term memory (LSTM) network is a particular type of RNN \cite{lstm_into} which can learn long-term dependence. The main difference between the LSTM and traditional RNN is that instead of using the input units and previous state units to update the state vector, it computes the output by implementing three gate units: forget gate unit, the input gate unit, and the output gate unit. These gate units control the flow of information from the input and current state to the next state. Similar to RNN, the associated weights of LSTMs are also optimized using BTT. Some major hyperparameters associated with LSTM are the number of hidden layers, number of hidden units in the layers, learning rate, dropout rate, batch size, etc.

%%%%%%%%%%%%%%%%%%%%%%%%%%%%%%

\subsection{Overview of Gaussian Bayesian Optimization}

Suppose $f(x)$ is a computationally expensive function lacking a known structure and does not have observable first or second-order derivatives, and as a result, finding solutions using Newton or quasi-Newton is not possible. Bayesian Optimization (BO) is a class of optimization algorithms focusing on solving the following problem.

\setlength{\abovedisplayskip}{3pt}
\begin{align}
    max_{x\epsilon A}f(x)\nonumber.
\end{align}

BO has two main components: an objective function ($f(x)$) and a Bayesian Statistical model, $p(x,f(x))$. The second component $p(x,f(x))$ is used for modeling the objective function, which describes potential values for $f(x)$ at a candidate point $x$ and an acquisition function for deciding where to sample next.

A Gaussian regression process is typically selected as the statistical model, which considers the prior distribution to be multivariate normal (MVN) with a particular mean vector and covariance matrix. At first, some initial steps are used to build the prior MVN density. Suppose, the set $k$ pairs of parameters and their corresponding objective function values after $k$ initial steps $D=$ $(x_1,f(x_1))$, $(x_2,f(x_2))$,$\ldots$,$(x_k,f(x_k))$. The mean vector of the prior MVN distribution is constructed by evaluating a mean function $\mu_0$ at each $x_i$. The covariance matrix, $\Sigma_0$, is constructed by evaluating a specific covariance function or kernel at each pair of points $(x_i,x_j)$. The kernel is chosen so that the points that are closer in the input space are more positively correlated, which means if $||x-x'||<||x-x''||$, then $\Sigma_0(x,x')>\Sigma_0(x,x'')$ where $||.||$ is some distance function. The usual choices for the kernel are the Gaussian or Matern kernel.

Expected improvement is used as the acquisition function \cite{human_problem}. After $n^{th}$ evaluation, the expected improvement at point $x$ can be written as

\begin{align}
    EI_n(x)=E_n\big[ (f(x)-f_n^*)^{+} \big] \nonumber
\end{align}

where, $f_n^*=max[f(x_1),f(x_2),...,f(x_n)]$ and $a^+=max(0,a)$. And the posterior density is updated as

\begin{align}
    f(x)|f(x_{1:n}) \sim MVN(\mu_n(x),\Sigma_n(x)) \nonumber
\end{align}

where, $\mu_n (x)=\mu_0(x)+\Sigma_0  (x,x_{1:n})\Sigma_0 (x_{1:n},x_{1:n})^{-1} \big( f(x_{1:n})-\mu_0(x_{1:n})\big)$ and $\Sigma_n(x)=\Sigma_0(x,x)-\Sigma_0 (x,x_{1:n})\Sigma_0 (x_{1:n},x_{1:n})^{-1}\Sigma_0 (x_{1:n},x)$. The next evaluation is done on the point where $EI_n(x)$ is maximum. The outline of BO for maximum number of iteration $N$ is described below

%For faster computation, and to make the solution numerically stable, a Cholesky decomposition is done and then the linear system of equations is solved \cite{bayesian_normal_theory}. 

\begin{itemize}

    \item  Initialize the prior $p(x,f(x))$.
    
    \item For every iteration $i \le N$,
    
    \begin{itemize}

    \item Obtain the new set of hyperparameter by solving
    \begin{align}
       \hat x = argmax_{x} EI_n(x). \nonumber
    \end{align}

    \item Evaluate the objective function at $\hat x$ as $f(\hat x)$.

    \item Update $p(x,f(x))$ by finding the posterior distribution using $D=D \cup \big(\hat x,f(\hat x)\big)$. 
    
    \end{itemize}
    
    \item Stop if $i=N$ or any other stopping criteria has been fulfilled.

\end{itemize}

\subsection{Data Splitting and forecasting strategy}
\label{data_theory}

For every station, we split the data set into three parts: training, validation, and test data set. The training data set is used for training a specific model, and the validation data set is used to monitor the performance of the trained model and find the best candidate model. One common approach while splitting the data set in this manner is to equal the length of the validation and test data set \cite{data_split_1,data_split_2}. Suppose $\mathbf{y}=(y_1$, $y_2$, $y_3$, ..., $y_T)$ be the time series data with $T$ observations, where $y_t$ indicates the observed data at $t^{th}$ timepoint. If $|Tr|$, $|V|$, and $|Te|$ be the length of training, validation, and test series, respectively, we can write the three series as follows.

\begin{itemize}
    \item Training: $\mathbf{y}_{Tr}=(y_1,...,y_{T-(|V|+|Te|)})$
    \item Validation: $\mathbf{y}_{V}=(y_{T-(|V|+|Te|-1)},...,y_{T-|Te|})$
    \item Test: $\mathbf{y}_{Te}=(y_{T-|Te|+1},...,y_{T})$
\end{itemize}

Note that, $\mathbf{y}_{Tr}$, $\mathbf{y}_{V}$ and $\mathbf{y}_{Te}$ contains the first $|Tr|$, the next $|V|$ and the last $|Te|$ observations of $\mathbf{y_{T}}$. Also, $|Tr|+|V|+|Te|=T$ and $|V|=|Te|$. Then we normalize $\mathbf{y}_{Tr}$, $\mathbf{y}_{V}$ using

\begin{equation}
    \mathbf{z}_{Tr}=\frac{\mathbf{y}_{Tr}- \min(\mathbf{y}_{Tr})}{\max(\mathbf{y}_{Tr}) - \min(\mathbf{y}_{Tr})}, \text{ and}
\end{equation}

\begin{equation}
    \mathbf{z}_{V}=\frac{\mathbf{y}_{V}- \min(\mathbf{y}_{Tr})}{\max(\mathbf{y}_{Tr}) - \min(\mathbf{y}_{Tr})}.
\end{equation}

For univariate time series forecasting problems, the inputs at a time point are simply the previous observations of that series. Suppose the number of lag values considered as input is $m$, then the input matrix $\mathbf{X}(m)$ and output vector $\mathbf{y}(m)$ can be expressed as

\begin{gather}
\label{input_matrix}
\mathbf{X}(m)
 =
  \begin{bmatrix}
    z_{1} & z_{2} & z_{3} & \ldots & z_{m} \\
    z_{2} & z_{3} & z_{4} & \ldots & z_{m+1} \\
\vdots & \vdots & \vdots & \ldots & \vdots \\
 z_{Tr-m} & z_{Tr-m} & z_{Tr-m+1} & \ldots & z_{Tr-1} 
   \end{bmatrix}, \text{ and}
\end{gather}

\begin{equation}
\label{output_vector}
      \mathbf{y}(m)=(z_{m+1},z_{m+1},\ldots,z_{Tr}).
\end{equation}

After training the model using $\mathbf{z}_{Tr}$, suppose we want to forecast for the next $|V|$ time points, say $\mathbf{\hat z}_{V}=\big (  \hat{z}_{T-(|V|+|Te|-1)}$, $\hat{z}_{T-(|V|+|Te|-2)}$, $...$, $\hat{z}_{T-|Te|)}\big )$. In our study, a multi-step forecasting scheme \cite{multi_step_defn} is considered. The input vector for forecasting at time point $|Tr|+h$ using a model that selected LAG as $m$, $\mathbf{x_{|Tr|+h}}$ can be expressed as the following.

\begin{equation}
        \mathbf{x}_{|Tr|+h}=\begin{cases}
        (z_{Tr}, \ldots, z_{|Tr|+m-1}), & \text{if $h=1$}\\
        (\hat z_{|Tr|+h-1}, \ldots, \hat z_{|Tr|+1},y_{|Tr|},...,z_{|Tr|+m+h-2}) & \text{if $1<h<=m$}\\
        (\hat z_{|Tr|+h-1}, \ldots, \hat z_{|Tr|+1},z_{|Tr|}) & \text{if $m>h$}\\
                  \end{cases}
\end{equation}

Here the input for the first forecast contains information from observed values, but the second and later forecast values will be dependent on both the observed values and previously forecasted values.

\subsection{Use Bayesian Optimization to find the optimal lag}
\label{proposed_lag}

The dimension of $\mathbf{X}(m)$ is $(Tr-m) \times m$ and the length of $\mathbf{y}(m)$ is $Tr-m$. Note that the values and dimensions of $\mathbf{X}(m)$ and $\mathbf{y}(m)$ depend on $m$. That is why $\mathbf{X}(m)$ and $\mathbf{y}(m)$ can be written as a function of $m$. We will use this relationship to find the optimal value of $m$ using Bayesian Optimization. Let $H_1$ be the set of hyperparameters associated with the structure of LSTM models. Our study also considers the number of selected lags as a hyperparameter. Then the set of hyperparameters we want to find the optimal value of any loss function is $H=(m, H1)$.

In general, the search space of BO is continuous. On the other hand, $m$ is discrete. Additionally, some other hyperparameters can be discrete. To mitigate this problem, instead of $H$, we use BO to optimize another set of hyperparameters $H2$ where every element of $H2$ takes continuous values. Before training an LSTM model, we convert $H2$ into $H$ by using 

\begin{equation}
    H=g(H2).
\end{equation}

Here the function $g(.)$ operates element-wise and rounds the evelemt only if it is discrete; else keeps the original value. Finally, we consider the negative of mean squared error as the objective function, which can be expressed as 

\begin{align}
        f(H2)=\displaystyle - \frac{\sum_{i=T-(|V|+|Te|-1)}^{T-|Te|}(z_i-\hat{z}_i)^2}{|V|}.  \label{obj_funct}
\end{align}

The negative sign is taken in \ref{obj_funct} because, by default, BO finds the maximum possible value, but any forecasting algorithm's objective is to get as less $RMSE$ as possible.

In a single iteration of BO, we do the following.

\begin{enumerate}
    \item Convert $H2$ into $H$ using $H=(m,H1)=g(H2)$.
    
    \item Find $\mathbf{X}(m)$ and $\mathbf{y}(m)$ is found from the training data set using equation \ref{input_matrix} and \ref{output_vector}. 
    
    \item Train a model using structural hyperparameter set $H1$. 
    
    \item Use the trained model has been used to find forecasts for the next $|V|$ timepoints,  
    
    \item After that, the value of the objective function $f(H)$ has been found for this particular set of hyperparameter ($H$). 

\end{enumerate}

\subsection{Train the final proposed model}
\label{train_final}

The model with the highest $f(H)$ value in the BO process only uses information from $\textbf{y}_{Tr}$. If we use this model as our finalized model, our final forecast may disregard the change in pattern that might occur in the last part ($\textbf{y}_{V}$. As a result, it is necessary to train the final model using information from training and validation data \cite{mape_alt}.

Suppose the optimized set of hyperparameters from BO is $H_{Opt}=(m_{Opt},H1_{Opt})$. For training the final model, we first normalize the training and validation data set ($\mathbf{y}_{Tr, V}$) to get $\mathbf{y}_{Tr, V}$. Then use $\mathbf{z}_{Tr,V}$ to get $\mathbf{X}(m_{Opt})$ and $\mathbf{y}(m_{Opt})$ via the data splitting process mentioned in subsection \ref{data_theory}. Finally, train a model using the hyperparameter set $H1_{Opt}$. After that, this finalized model forecasts for the next $|Te|$ time points. Suppose the forecasts using this model is $\hat{\mathbf{z}} _{Te}$. Finally, we get the final forecast by denormalizing $\hat{\mathbf{z}} _{Te}$ using

\begin{equation}
    \hat{\mathbf{y}} _{Te}=\hat{\mathbf{z}} _{Te} \times \big ( \max(\mathbf{y}_{Tr,V})-\min(\mathbf{y}_{Tr,V}) \big ) + \min(\mathbf{y}_{Tr,V})
\end{equation}

\begin{figure}[!ht]
	\centering	
	\includegraphics[scale=1]{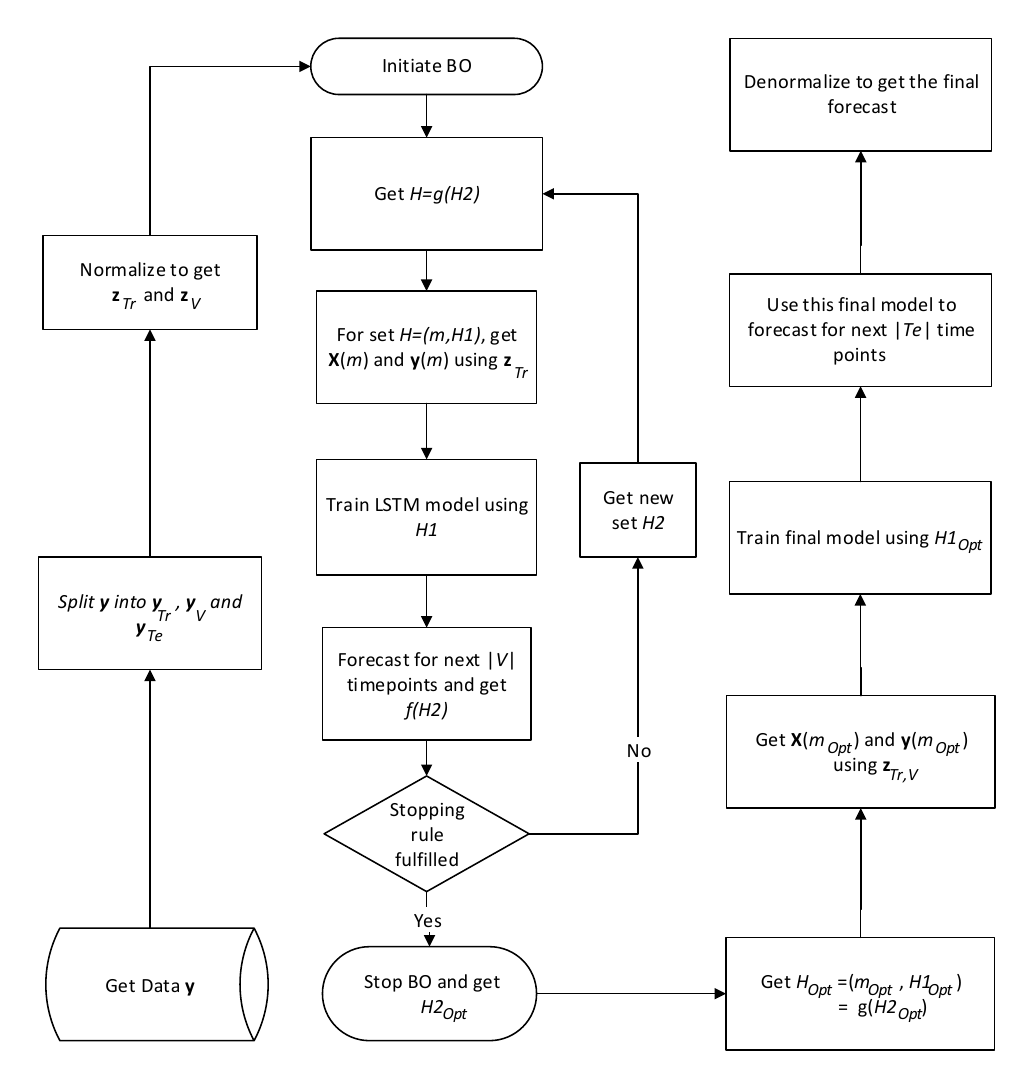}
	\caption{Overview of the whole process}
	\label{Full_process}
\end{figure}

The entire process of finding optimal $m$ and training our proposed model is given in Figure \ref{Full_process}.

\subsection{Evaluation criteria for point forecast}
\label{point_metrics}

To evaluate the performance of point forecasts, we use four measures. According to all these four measures, a lower value implies that a model performs better than a model having a higher value. Suppose $y_1,y_2,y_3,\ldots,y_h$ be the observed values and if $\hat{y}_1, \hat{y}_2, \hat{y}_3,\ldots,\hat{y}_H$ are the point forecasts on the same time points.

Then the first two measures, root mean squared error ($RMSE$) and mean absolute error ($MAE$), can be expressed as

\begin{equation}
\label{rmse}
            RMSE=\displaystyle \sqrt{\frac{\sum_{i=1}^h(y_i-\hat{y}_i)^2}{h}},
\end{equation}

and

\begin{equation}
\label{mae}
            MAE=\displaystyle \frac{\sum_{i=1}^h(|y_i-\hat{y}_i|)}{h}.
\end{equation}

Here $h$ is the length of the forecasting horizon. Additionally, to evaluate the performance of the forecasts relative to the actual values, we use symmetric mean absolute percentage error ($SMAPE$) that can be defined as:

\begin{equation}
\label{smape}
            SMAPE=\displaystyle \frac{100 \%}{h} \sqrt{\sum_{i=1}^h\frac{|y_i-\hat{y}_i|}{(|y_i|+|\hat{y_i}|)/2}}
\end{equation}

 However, $SMAPE$ is unstable with values close to zero \cite{mape_problem}. In this regard, Suilin \cite{mape_alt} proposed to replace the denominator of \ref{smape} by the following:

\begin{equation}
\label{smape_alter}
            max(|y_i|+|\hat{y_i}|+\epsilon, 0.5+\epsilon)
\end{equation}

This alternate measure skips dividing by values close to zero by replacing those with an alternative positive constant. We set the hyperparameter $\epsilon=0.1$, suggested by Suilin \cite{mape_alt}.

Finally, we use average rank ($ARank$). If $\hat y_{i,m}$ be the forecast for timepoint $i$ by model $m$, then $ARank$ can be defined as

\begin{equation}
    ARank= \sum_{i=1}^h \frac{R_{i,m}}{h},
\end{equation}

where $R_{i,m}$ is the rank of forecast of time point $i$ by model $m$ among the $M$ candidate model in ascending order. The lowest and highest possible value of $ARank$ is $1$ and $M$, respectively.

%%%%%%%%%%%%%%%%%%%%%%%%%%%%%%%%%%%%%%%%%%%%%%%%%%%%%%%%%%%%%%%%%%%%%%%%%%%%%%%

\subsection{Comparing the performance of multiple methods}

\label{testing_two_step}

According to any standard measure, if the performance of a model is found to be better than all other candidate models, then it is easy to conclude that the method is better. However, if no model is found superior in every case, then it is necessary to conduct statistical tests to compare their performances \cite{test_needed}. The two-step testing procedure proposed by Demsar \cite{multiple_machine_learning} can be used in this regard. This method was primarily developed to compare classification methods over multiple data sets. Later, Abbasimehra et al. \cite{extended} implemented this method for comparing multiple forecasting models over multiple data sets. For this, according to that particular measure, rank has been found for different models and datasets, and the average rank for each method is calculated. Then for the first testing step, a hypothesis test is performed to check whether there is any significant difference in the performance of any two proposed models. If the result is significant, pairwise post hoc tests have been performed.

Suppose we have applied $k$ algorithms on $N$ total data sets, and according to any measure, $r_i^j$ is the rank of the $j^{th}$  algorithms on the $i^{th}$ data set. For the first step, a Friedman statistic has been compute using
\begin{align*}
\chi_F^2=\displaystyle \frac{12N}{k(k+1)}\Big[ \sum_{j}R_j^2 - \displaystyle \frac{k(k+1)^2}{4} \Big], 
\end{align*}

where $R_j=\sum_{i=1}^N r_i^j$ is the average rank of $j^{th}$ algorithm. Under the null hypothesis ($H_0$) that all the models perform identical $\chi_F^2$ follows a $\chi^2$ distribution with $(k-1)$ degrees of freedom ($df$). However, Iman and Davenport \cite{friedman_alt} showed that Friedman's $\chi_F^2$ is undesirably conservative and proposed a different test statistic that can be stated as

\begin{align}
F_F=\displaystyle \frac{(N-1)\chi_F^2}{N(k-1)-\chi_F^2}
\end{align}

which follows as a $F-$distribution with $df$ $(k-1)$ and $(k-1)(N-1)$ under $H_0$. If $H_0$ is rejected in the first step, only then the second-stage testing is performed, where Hochberg's post hoc test is used to check whether the reference model performs better than the other candidate models using the statistic 

\begin{align}
z_{Model}=\displaystyle \frac{R_{Model}-R_{Reference}}{\sqrt{ \displaystyle \frac{k(k+1)}{6N}}}.
\end{align}

Here the reference model is our proposed model. The critical values of $z_{Model}$ can be found in a standard normal distribution table. For Hochberg's test, the p-values of the $z$ values for $(k-1)$ comparisons with the control model are first computed and sorted from the smallest to the largest. Finally, the $i^{th}$ model is called significantly different from the reference model if the $p-value$ is smaller than $\frac{i}{k}\alpha$.

%%%%%%%%%%%%%%%%%%%%%%%%%%%%%	
\section{Case study}

\subsection{Selected models for comparison}
For comparing the performance of the proposed model ($PROP$) for different data sets we consider the following models.

\begin{enumerate}
    \item LSTM with number of past observations= seasonal length ($LSL$).
    \item LSTM with number of past observations= forecasting horizon ($LFH1$).
    \item LSTM with number of past observations= seasonal length $\times 1.25$ ($LFH1p25$).
    \item Support vector regression ($SVR$).
    \item Random Forest ($RF$).
    \item Seasonal autoregressive moving average ($SARIMA$).
    \item Exponential smoothing ($ETS$).
    
\end{enumerate}

For the model $1-3$, similar strategy as $PROP$ has been considered except the number of lag values has been before training the models and other hypperparameters has been optimized using BO. Additionally, because $RF$ and $SVR$ are not too computationally extensive, we use grid search to find the best set of hyperparameters for them \cite{grid_search_rf,grid_search_svr}. In this case, the similar data-splitting process mentioned in subsection \ref{data_theory} is used.

\subsection{Libraries used}

To implement our proposed method, we used \texttt{Python} library \texttt{TensorFlow} to train LSTM models and \texttt{bayesopt} for implementing BO. Additionally, to fit the two machine learning models: RF and SVR models, \texttt{Python} libraries \texttt{sklearn} has been used. On the other hand, to fit the two statistical forecasting models: SARIMA and ETS, the package \texttt{forecast} programming language \texttt{R} has been used.

\subsection{Data and preprocessing}

The data sets are taken from Bangladesh Meteorological Department. Out of the 35 weather stations,  in this research, rainfall data of nine selected stations those are located in the nine divisional districts named as: Barisal (BR), Chittagong (CH), Comilla (CM), Dhaka (DH), Khulna (KH), Mymensingh (MY), Rajshahi (RJ), Rangpur (RN) and Sylhet (SY). The starting and ending months are given in table \ref{data_info}. The data sets contain no missing values. Then every data set is split into 3 parts: training, validation and test. For every cases, we keep the length of the validation and test series fixed as $60$ which means $|V|=|Te|=60$.

\begin{table}[!ht]
\centering
\caption{Starting and ending months and the number of missing values for the Rainfall data of 9 stations}
\label{data_info}
\begin{tabular}{cccccccccc}
\hline
      & BR      & CH      & CM      & DH      & KH      & MY      & RJ      & RN      & SY      \\ \cline{2-10} 
Start & 1948-01 & 1949-01 & 1948-01 & 1953-01 & 1948-01 & 1948-01 & 1964-01 & 1954-01 & 1956-01 \\
End   & 2020-07 & 2020-10 & 2020-08 & 2020-10 & 2020-08 & 2020-08 & 2020-09 & 2020-07 & 2020-07 \\ \hline
\end{tabular}
\end{table}

\begin{figure}[!ht]
	\centering	
	\includegraphics[scale=.65]{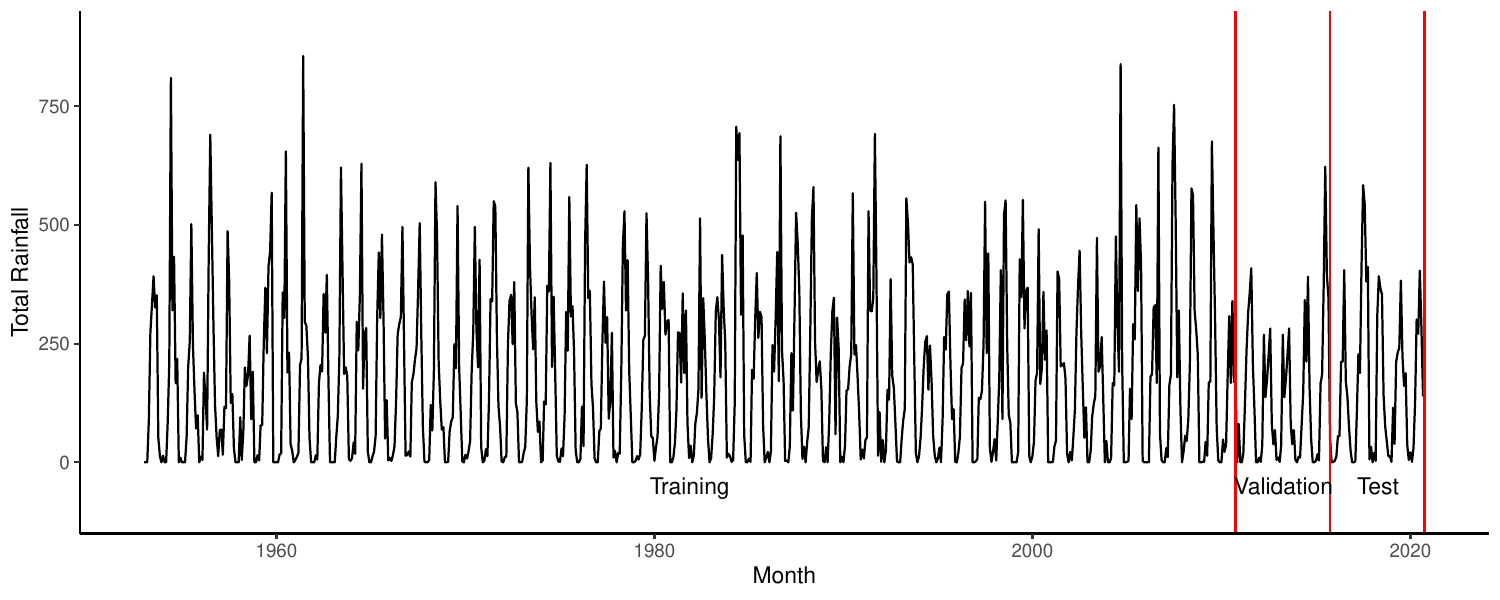}
	\caption{Data splitting process for station DH}
	\label{Main_Series_Dhaka}
\end{figure}

To check the stationarity of our series, we first perform an augmented Dicky-Fuller (ADF) test \cite{adf}. Since none of the $p-values$ are bigger than $0.05$, we reject the null hypothesis stating that the data is non stationary for every series. Then because neural network models cannot model seasonality perfectly \cite{data_preprocess}, we first deseasonalize the series using STL decomposition and then proceed with the deseasonalized series for training using $PROP$ modeld. Note that, the process of testing nonstationarity and deseasonalizing the data is performed only on the training and validation data sets; we keep the test data sets intake because we do not want the performance on the actual data sets, not on the transformed ones.

\begin{table}[!ht]
\centering
\caption{Statistic value from ADF test}
\begin{tabular}{lrrrrrrrrr}
\hline
          & BR      & CH      & CM      & DH      & KH      & MY      & RJ      & RN      & SY      \\ \cline{2-10} 
$Statistic$ & -12.479 & -16.696 & -11.139 & -10.787 & -13.043 & -11.662 & -16.073 & -11.729 & -11.922 \\
$p-value$   & $0.010$   & $0.010$   & $0.010$   & $0.010$   & $0.010$   & $0.010$   & $0.010$   & $0.010$   & $0.010$   \\ \hline
\end{tabular}
\end{table}

%%%%%%%%%%%%%%%%%%%%%%%%%%%%%%%%%%%%%%%%%%%%%%%%%%%%%%%%%%%%%%%%%%%%%%%%

\subsection{Results}
\label{point_forecast_results}

In our experiment, apart from the number of lag values as input ($m$), we are considering five LSTM hyperparameters:  dropout rate ($dr$), learning rate($lr$), number of hidden units in hidden layer 1 ($hu_1$), number of hidden units in layer 2 ($hu_2$) and batch size ($b$).Hence the set of hyperparameters ($H_1$) we want to optimize using BO has six items. The selected lag values for $PROP$ are given in Table \ref{selected_lags}.

\begin{table}[!ht]
\centering
\caption{Selected Lags using for $PROP$}
\begin{tabular}{ccccccccc}

\hline
BR      & CH      & CM      & DH      & KH      & MY      & RJ      & RN      & SY \\ \hline
34      & 32         & 36      & 45    & 34     & 38         & 31       & 33      & 35     \\ \hline
\end{tabular}
\label{selected_lags}
\end{table}

\begin{figure}[!ht]
	\centering	
	\includegraphics[scale=.75]{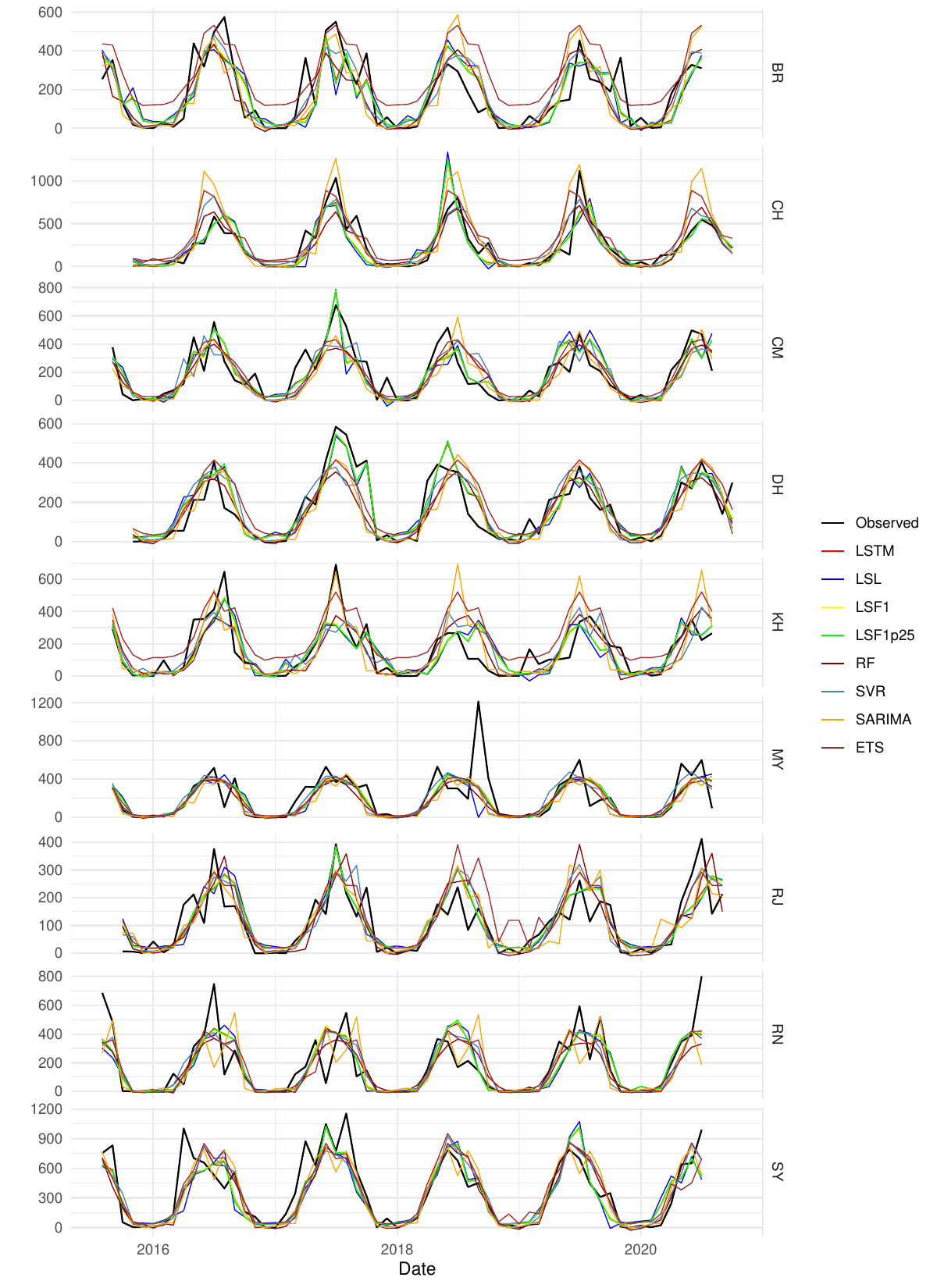}
	\caption{Observed and forecasted monthly rainfall by all candidate methods for all stations.}
	\label{all_forecasts}
\end{figure}

The performance of $PROP$ and the other selected models are given in Table \ref{performance_all}. Among the four LSTM considered, $PROP$ performs better than the other three LSTM models since it has the lowest $RMSE$, $MAE$, $SMAPE$, and $ARank$ values among them for all stations. On the other hand, $LSL$ constantly performs worse than the other three LSTM models, regardless of what measure we use for evaluating the performance. This provides evidence that the selected lag for for $LSL$, $12$ is insufficient for training an LSTM. The comparison between $LFH1$ and $LFH1p25$ is not consistent. $LFH1p25$ has lower $RMSE$ and $MAE$ for station RN; and has lower $SMAPE$ for station CH. For all other stations, $LFH1$ performs better than $LFH1p25$.

\setlength{\tabcolsep}{2.9pt}
\begin{table}[!ht]
\renewcommand{\arraystretch}{.75}
\centering
\caption{Performance of forecasts from all candidate models.}
\label{performance_all}

\begin{tabular}{c|c|ccccccccc}
\hline
\multirow{2}{*}{Measure} & \multicolumn{1}{c|}{\multirow{2}{*}{Model}} & \multicolumn{9}{c}{Station}                                                                                                                                                                                                    \\ \cline{3-11} 
                         & \multicolumn{1}{c|}{}                       & \multicolumn{1}{r}{BR} & \multicolumn{1}{r}{CH} & \multicolumn{1}{r}{CM} & \multicolumn{1}{r}{DH} & \multicolumn{1}{r}{KH} & \multicolumn{1}{r}{MY} & \multicolumn{1}{r}{RJ} & \multicolumn{1}{r}{RN} & \multicolumn{1}{r}{SY} \\ \hline
\multirow{8}{*}{RMSE}    & LSTM                                        & 18.270                 & 12.679                 & \textbf{14.677}        & \textbf{11.999}        & \textbf{13.966}        & \textbf{12.952}        & \textbf{15.497}        & \textbf{15.884}        & \textbf{14.696}        \\
                         & LSL                                         & 21.792                 & 15.467                 & 17.598                 & 14.234                 & 16.193                 & 16.655                 & 18.002                 & 18.148                 & 18.035                 \\
                         & LFH1                                        & 18.720                 & 12.929                 & 14.788                 & 12.024                 & 14.058                 & 13.165                 & 15.895                 & 16.257                 & 14.810                 \\
                         & LFH1p25                                      & 18.858                 & 13.164                 & 14.977                 & 12.474                 & 14.354                 & 13.254                 & 16.045                 & 16.197                 & 14.969                 \\
                         & SVR                                         & 19.142                 & \textbf{12.061}        & 16.030                 & 15.595                 & 15.194                 & 14.170                 & 18.977                 & 16.498                 & 14.805                 \\
                         & RF                                          & \textbf{17.662}        & 12.583                 & 19.173                 & 16.852                 & 16.818                 & 13.195                 & 17.233                 & 16.004                 & 14.972                 \\
                         & SARIMA                                      & 21.972                 & 20.163                 & 16.869                 & 17.848                 & 19.688                 & 15.402                 & 17.768                 & 22.214                 & 15.992                 \\
                         & ETS                                         & 25.721                 & 16.555                 & 15.123                 & 16.474                 & 20.379                 & 13.575                 & 18.228                 & 17.452                 & 14.833                 \\ \hline
\multirow{8}{*}{MAE}     & LSTM                                        & 12.884                 & \textbf{7.528}         & \textbf{11.161}        & \textbf{9.176}         & \textbf{9.610}         & \textbf{6.983}         & \textbf{11.346}        & \textbf{9.923}         & 9.739		            \\
                         & LSL                                         & 15.519                 & 9.116                  & 13.179                 & 10.934                 & 11.448                 & 8.491                  & 13.296                 & 11.594                 & 11.880                 \\
                         & LFH1                                        & 13.133                 & 7.654                  & 11.223                 & 9.248                  & 9.666                  & 7.042                  & 11.584                 & 10.115                 & 9.836                  \\
                         & LFH1p25                                      & 13.210                 & 7.765                  & 11.422                 & 9.429                  & 9.893                  & 7.234                  & 11.662                 & 10.111                 & 9.949                  \\
                         & SVR                                         & 13.006                 & 7.924                  & 12.171                 & 10.544                 & 10.324                 & 7.757                  & 13.278                 & 10.471                 & \textbf{9.675}         \\
                         & RF                                          & \textbf{11.911}        & 8.243                  & 14.802                 & 11.503                 & 11.875                 & 7.489                  & 12.608                 & 10.077                 & 9.823                  \\
                         & SARIMA                                      & 14.232                 & 11.455                 & 12.183                 & 12.351                 & 12.578                 & 8.932                  & 13.906                 & 12.668                 & 10.093                 \\
                         & ETS                                         & 22.774                 & 11.403                 & 11.862                 & 12.761                 & 17.159                 & 7.552                  & 14.309                 & 11.567                 & 10.938                 \\ \hline
\multirow{8}{*}{SMAPE}   & LSTM                                        & 14.765                 & \textbf{8.950}         & \textbf{13.203}        & \textbf{11.227}        & \textbf{12.870}        & \textbf{9.563}         & \textbf{12.917}        & \textbf{11.420}        & 10.489			        \\
                         & LSL                                         & 18.025                 & 10.923                 & 15.489                 & 13.704                 & 15.454                 & 11.864                 & 15.426                 & 13.363                 & 13.108                 \\
                         & LFH1                                        & 15.118                 & 9.021                  & 13.268                 & 11.291                 & 12.982                 & 9.642                  & 13.227                 & 11.601                 & 10.605                 \\
                         & LFH1p25                                      & 15.145                 & 9.198                  & 13.549                 & 11.497                 & 13.239                 & 9.956                  & 13.238                 & 11.611                 & 10.756                 \\
                         & SVR                                         & \textbf{14.253}        & 9.880                  & 14.629                 & 12.132                 & 13.120                 & 10.746                 & 14.201                 & 12.284                 & \textbf{9.867}         \\
                         & RF                                          & 13.348                 & 10.032                 & 17.173                 & 13.302                 & 15.279                 & 10.322                 & 14.093                 & 11.688                 & 10.049                 \\
                         & SARIMA                                      & 15.143                 & 11.678                 & 14.680                 & 14.390                 & 15.366                 & 12.780                 & 17.004                 & 14.341                 & 10.530                 \\
                         & ETS                                         & 26.688                 & 13.838                 & 14.519                 & 14.946                 & 21.855                 & 10.513                 & 16.277                 & 13.677                 & 11.567                 \\ \hline
\multirow{8}{*}{ARank}   & LSTM                                        & \textbf{3.683}         & \textbf{3.300}         & \textbf{3.950}         & \textbf{3.717}         & \textbf{3.367}         & \textbf{3.383}         & \textbf{3.450}         & \textbf{3.317}         & \textbf{3.933}         \\
                         & LSL                                         & 6.000                  & 5.550                  & 5.967                  & 6.183                  & 5.833                  & 5.883                  & 5.900                  & 5.717                  & 6.383                  \\
                         & LFH1                                        & 4.150                  & 3.667                  & 4.033                  & 4.167                  & 3.617                  & 3.617                  & 4.083                  & 3.883                  & 4.117                  \\
                         & LFH1p25                                      & 3.950                  & 4.100                  & 4.483                  & 4.183                  & 4.067                  & 4.050                  & 4.333                  & 4.117                  & 4.467                  \\
                         & SVR                                         & 3.833                  & 4.483                  & 4.200                  & 3.817                  & 4.100                  & 4.950                  & 4.417                  & 4.717                  & 4.150                  \\
                         & RF                                          & 3.733                  & 4.567                  & 4.450                  & 4.300                  & 4.750                  & 4.333                  & 4.267                  & 3.917                  & 4.150                  \\
                         & SARIMA                                      & 4.200                  & 4.550                  & 4.233                  & 4.467                  & 4.183                  & 5.183                  & 4.817                  & 4.733                  & 3.983                  \\
                         & ETS                                         & 6.450                  & 5.783                  & 4.683                  & 5.167                  & 6.083                  & 4.600                  & 4.733                  & 5.600                  & 4.817                  \\ \hline
\end{tabular}

\end{table}

Additionally, we can easily conclude that $PROP$ performs better than $SARIMA$ and $ETS$ since it has lower $RMSE$, $MAE$, $MAE$ and $ARank$ values for all stations. In case if we consider $ARank$ as our metric, $PROP$ performs better than the other for all stations. However, the comparison between $RF$ and $SVR$ is more complex. $RF$ has the lowest $RMSE$ and $MAE$ for one station each. On the other hand, $SVR$ has the lowest $RMSE$, $MAE$, and $SMAPE$ for one, one, and two stations, respectively. As a result, to check whether $PROP$ performs better than these two models, using the two-step procedure mentioned in the subsection \ref{testing_two_step} is necessary.

To complete Desmar's two-step test, we first rank the models based on the $RMSE$ in ascending orders for every station and then calculate the average rank for every station; and finally, use these average ranks for performing the procedure mentioned in section \ref{testing_two_step}. From Table \ref{tab:two_step_test}, the $(F_F, p-value)$ pairs for testing whether there is any difference in the performance of the models are $(13.430,0.000)$, $(21.518, 0.000)$ and $(15.341,0.000)$ for $RMSE$, $MAE$ and $SMAPE$ respectively. Since all the corresponding $p-values$ are very small, we conclude that the performance of at least two models is not identical, and we proceed to the second stage of Desmar's two-step test.

\begin{table}[!ht]
\centering
\caption{Two-step testing for comparing the performances of $PROP$ with other candidate models.}
\label{tab:two_step_test}

\begin{tabular}{c|c|l|rrrrrrr}
\hline
\multirow{2}{*}{Measure} & \multirow{2}{*}{} & \multicolumn{1}{c|}{Step 1} & \multicolumn{7}{c}{Step 2}                                                                                                                                                                                                          \\ \cline{3-10} 
                         &                   & \multicolumn{1}{c|}{$F_F$}  & \multicolumn{1}{c}{$z_{LSL}$} & \multicolumn{1}{c}{$z_{LFH1}$} & \multicolumn{1}{c}{$z_{LFH1p25}$} & \multicolumn{1}{c}{$z_{SVR}$} & \multicolumn{1}{c}{$z_{RF}$} & \multicolumn{1}{c}{$z_{SARIMA}$} & \multicolumn{1}{c}{$z_{ETS}$} \\ \hline
\multirow{2}{*}{$RMSE$}  & $Statistic$       & 13.430                      & 4.330                         & 1.155                          & 2.021                            & 2.791                         & 2.598                        & 4.907                            & 4.138                         \\
                         & $p-value$         & 0.000                       & 0.000                         & 0.248                          & 0.043                            & 0.005                         & 0.009                        & 0.000                            & 0.000                         \\ \hline
\multirow{2}{*}{$MAE$}   & $Statistic$       & 21.518                      & 4.523                         & 1.251                          & 1.925                            & 2.502                         & 2.694                        & 5.004                            & 4.811                         \\
                         & $p-value$         & 0.000                       & 0.000                         & 0.211                          & 0.054                            & 0.012                         & 0.007                        & 0.000                            & 0.000                         \\ \hline
\multirow{2}{*}{$SMPE$}  & $Statistic$       & 15.341                      & 4.426                         & 0.866                          & 1.925                            & 2.021                         & 2.309                        & 4.234                            & 4.619                         \\
                         & $p-value$         & 0.000                       & 0.000                         & 0.386                          & 0.054                            & 0.043                         & 0.021                        & 0.000                            & 0.000                         \\ \hline

\end{tabular}
\end{table}

In the second step, we consider $10\%$ significance level. In Table \ref{tab:two_step_test}, we see that all $z-values$ are positive for every model across all measures, which is analogous to the fact that $PROP$ has a lower average rank according to $RMSE$, $MAE$, and $SMAPE$. 

Here we have seven models for comparison. We say $PROP$ performs better than the model with $i^{th}$ smallest $p-value$ if it is smaller that $0.10 \frac{i}{7}$. We conclude that we can conclude that $PROP$ performs better than all models except $LFH1$. It should be noted that, according to all the measures $PROP$ performs better that $LFH1$ yet we cannot reject the null hypothesis here. This happens because $LFH1$ perform better than the other candidate models except $PROP$, and as a result, it has average ranks smaller than others. Additionally, the performance of the other models can be ranked using the $p-values$ in the second step from table \ref{tab:two_step_test}. The models with higher $p-values$ indicate that the model is performing close to $PROP$ as they are tougher to reject, and it can be said that they are performing better than the models having smaller $p-values$. For example, according to $RMSE$, $LFH1$ performs better than $LFH1p25$. In general, according to all three measures, $LFH1$ performs better than all other models, and $ETS$ performs the worst. Additionally, the two machine learning models $RF$ and $SVR$ perform better than the two statistical methods ($ETS$ and $SARIMA$), but they got outperformed by $LFH1$ and $LFH1p25$. We have not completed the two-step test for $ARank$ since $PROP$ has lower values than all other candidate models across all stations.

%%%%%%%%%%%%%%%%%%%%%%%%%%%%%%%%%%	
\section{Conclusion}

In this study, we used BO to find the optimal lag inputs of $LSTM$ for forecasting monthly average rainfall of nine weather stations of Bangladesh. Then we compared models developed by our proposed model with some other models and showed that our model outperforms those methods in most of the cases.

Among the limitations of the study, the most obvious one is that rainfall is not solely dependent on past observations; rather, it depends on many factors like air quality, latitude, elevation, topography, seasons, distance from the sea, and coastal sea-surface temperature and many other factors. Moreover, the study has been done only for the stations located in the divisional cities of Bangladesh rather than considering all the stations of Bangladesh because the data for other stations were not available. 

The second major drawback is that we used a Gaussian Bayesian Optimization for finding the optimal set of hyperpaprameters of $LSTM$ that naturally searches for continuous hyperparametes. LAGS and some other hyperparameters are discrete on the other hand. As a result, during training our proposed model, we lost some information because of using the conditional rounding function. 

For potential future extensions, a study that uses other variables like temperature, location information, air quality can improve forecasting accuracy. Recently, parzen tree based Bayesian optimization algorithms has been popular as they have the capability to consider discrete space for hyperparameters. So we can suspect that using Parzen Tree based algorithms might further improve the proposed method.

\bibliographystyle{unsrtnat}

\bibliography{Manuscript_TwoStage}

\end{document}